\documentclass{article}
\usepackage{epsfig}
\usepackage{amsfonts}
\usepackage{amsmath}
\usepackage{color}
\usepackage{epstopdf}

\begin{document}

\title{Quantum computation using arrays of $N$ polar molecules in pendular states}

\date{}

\author{Qi Wei\thanks{State Key Laboratory of Precision Spectroscopy, East China Normal University, Shanghai 200062, China}\and Yudong Cao\thanks{Department of Computer Science, Purdue University, West Lafayette, IN 47907, USA}\and Sabre Kais\thanks{corresponding author: kais@purdue.edu}\textsuperscript{   \makebox[0.02in]{},\makebox[0.02in]{}}\thanks{Department of Chemistry and Physics, Purdue University, West Lafayette, IN 47907, USA; Qatar Environment and Energy Research Institute, HBKU, Qatar Foundation, Doha, Qatar}\and Bretislav Friedrich\thanks{Fritz-Haber-Institut der Max-Planck-Gesellschaft, Faradayweg 4-6, D-14195 Berlin, Germany}\and Dudley Herschbach\thanks{Department of Chemistry and Chemical Biology, Harvard University, Cambridge, Massachusetts 02138, USA}}

\maketitle

\begin{abstract}

We investigate several aspects of realizing quantum computation using entangled polar molecules in pendular states. Quantum algorithms typically start from a product state $|00\cdots 0\rangle$ and we show that up to a negligible error, the ground states of polar molecule arrays can be considered as the unentangled qubit basis state $|00\cdots 0\rangle$. This state can be prepared by simply allowing the system to reach thermal equilibrium at low temperature ($<1$ mK). We also evaluate entanglement, characterized by the concurrence of pendular state qubits in dipole arrays as governed by the external electric field, dipole-dipole coupling and number $N$ of molecules in the array. In the parameter regime that we consider for quantum computing, we find that qubit entanglement is modest, typically no greater than $10^{-4}$, confirming the negligible entanglement in the ground state. We discuss methods for realizing quantum computation in the gate model, measurement based model, instantaneous quantum polynomial time circuits and the adiabatic model using polar molecules in pendular states. 

\end{abstract}

\section{Introduction}

Quantum computers take advantage of superposition and entanglement to perform computations in ways that are beyond the reach of classical computers \cite{Bennett,Deutsch,Feynman,Shor,Grover,Chuang}. Among the many possible schemes for realizing quantum computers, arrays of trapped ultracold polar molecules subject to an external electric field are considered a promising approach \cite{Demille,andre,yelin,carr,Book2009,Friedrich,Lee,Kotochigova,Micheli,Charron,kuz,ni,lics,YelinDeMille,Wei1,Wei2, Wei3, Zhu,Felipe,Friedrich3}. In such a dipole array, each polar molecule acts as a qubit entangled with the other molecules via electric dipole-dipole interaction.  Using the Stark effect due to an inhomogeneous external electric field, qubits encoded in rotational states can be individually  addressed and manipulated. Such a system is scalable to large networks of coupled qubits. 

For the simplest case of a $^1\Sigma$ diatomic molecule, due to the Stark effect from external electric fields, the qubit eigenstates are linear combinations of spherical harmonics, with coefficients that depend markedly on the field strength. These are appropriately termed {\it pendular} states ~\cite{Friedrich2}, or field-dressed states \cite{Ticknor}. Quantum computation and quantum information processing are inevitably based on those pendular states. For such states,  the rotational spectrum and the dipole-dipole interaction differ qualitatively from those in pure rotational  states. 

In our previous work, we focused on a small system with only two polar molecules in pendular states \cite{Wei2, Zhu}. We studied entanglement measured by pairwise concurrence as a function of molecular dipole moment, rotational constant, strength of external field  and dipole-dipole coupling. We also evaluated a key frequency shift, $\triangle\omega$, induced by the dipole-dipole interaction,  which is essential for quantum logic gate operations \cite{Wei2}. For a given frequency shift, $\triangle\omega$, we numerically implemented NOT, Hadamard and CNOT gates on two qubits encoded in pendular states of polar molecules \cite{Zhu}. 

Here, we extend the system from two to $N$ qubits and examine the feasibility of quantum computation with  1-dimensional and 2-dimensional arrays of trapped dipoles in pendular states. First, we study the initialization of the the system and calculate the probability for the ground state of a linear dipole array with $N$ polar molecules to be in a pure qubit basis state $|000...0\rangle$, which is the desired initial state for most quantum algorithms \cite{Book}. Second, we study the complexity of different dipole arrays by calculating pairwise entanglement between any two polar molecules in order to check whether systems can be simplified by considering only nearest-neighbor interactions. Finally, we discuss the feasibility of realizing four different types of quantum computation based on arrays of polar molecules in pendular states, namely the the gate model \cite{Deutsch89,NC11}, measurement-based model \cite{RB01,BR01,RBB03}, instantaneous quantum polynomial-time circuits \cite{Bremner459,Shepherdrspa.2008.0443} and the adiabatic model \cite{FGGS00,kadowaki1998quantum}.

\section{Hamiltonian for arrays of polar molecules}

The Hamiltonian for $N$ identical trapped polar molecules subject to an external electric field takes the form
\begin{equation}
{\bf H}=\sum_{i=1}^{N}\left [\frac{p_i^2}{2m} + V_\text{trap}({\bf r}_i)
+ BJ_i^2 - {\boldsymbol \mu}\cdot{\boldsymbol \varepsilon}
\right ] + \sum_{i=1}^{N-1}\sum_{j=i+1}^{N}V_{d-d}^{i,j}
\end{equation}
where molecule $i$, with mass $m$, rotational constant $B$ and body-fixed dipole moment $\mu$, has translational kinetic energy $\frac{p_i^2}{2m}$, potential energy $V_\text{trap}$ within the trapping field, and rotational energy $BJ_i^2$ as well as interaction energy ${\boldsymbol \mu}\cdot{\boldsymbol \varepsilon}$ with the external field ${\bf \varepsilon}$ and dipole-dipole interaction energy $V_{d-d}^{i,j}$ with other molecules $j$ in the array. Although the external field strength ${\bf \varepsilon}$ differs slightly at the site of each dipole in order to provide addressability, for our purposes we neglect this variation. For a harmonic translational motion of the molecule inside the trapping well, the total energy $\frac{p_i^2}{2m} + V_\text{trap}({\bf r}_i)$ is nearly constant and thus can be removed from the Hamiltonian.

\begin{figure}[ht]
\begin{center}
\includegraphics[width=0.7\textwidth]{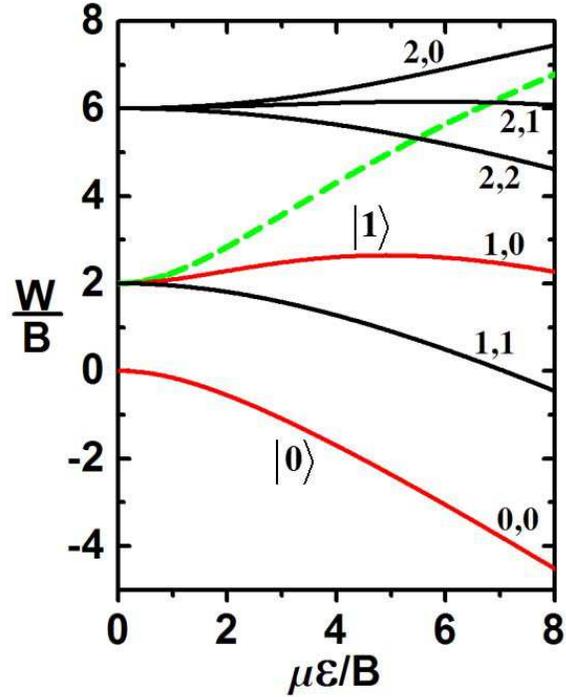}
\end{center}
\caption{Stark states eigenenergies for a polar diatomic molecule in a $^1\Sigma$ electronic state \cite{Hughes}, 
as functions of $\mu{\bf \varepsilon}/B$, with $\bf \mu$ the permanent dipole moment, ${\bf \varepsilon}$ the field strength, 
$B$ the rotational constant. States used as qubits (red curves) are labeled $|0\rangle$ and $|1\rangle$. In the field-free limit,
$|0\rangle$ correlates with the $J = 0$, $M_J = 0$ and $|1\rangle$ with the $J = 1$, $M_J = 0$ rotational states. Dashed curve (green) 
shows energy for transition between qubit states, $\triangle W = W^{(0)} - W^{(1)}$. If evaluated using customary units, the unitless ratio  
$\mu{\bf \varepsilon}/B$ is given by 0.0168 d(Debye)${\bf \varepsilon}(kV/cm)/B(cm^{-1})$} \label{Figure1}
\end{figure}

The terms $BJ_i^2 - {\boldsymbol \mu}\cdot{\boldsymbol \varepsilon}$ pertain to molecular rotational states that are strongly affected by the Stark effect interaction with the external field, which mixes the field-free rotational states \cite{Hughes}. Figure \ref{Figure1} shows the Stark eigenenergies for the lowest states of a $^1\Sigma$ diatomic polar molecule. The molecular qubits, $|0\rangle$ and $|1\rangle$, are defined as the indicated Stark states, $|J=0, M=0\rangle$ and $|J=1, M=0\rangle$ respectively. Actually the qubit states are appropriately termed ``pendular" \cite{Friedrich2}, since they arise due to a $cosine$ potential and the dipole orientations have broad angular ranges:
\begin{equation}
|0\rangle=\sum_{i=1}c_iY_i^0(\theta,\varphi),\;\;\;\;\;\;\;\;\;\;
|1\rangle=\sum_{i=1}c_i'Y_i^0(\theta,\varphi).
\end{equation}

Since the field strength is fixed for the processes of interest, the Hamiltonian can be recast as,
\begin{equation}
{\bf H}=\sum_{i=1}^{N}H_S^i  +
\sum_{i=1}^{N-1}\sum_{j=i+1}^{N}V_{d-d}^{i,j}
\end{equation}
where $H_S^i = BJ_i^2 - {\boldsymbol \mu}\cdot {\boldsymbol \varepsilon}$ is the Hamiltonian corresponding to the pendular states. In particular,
\begin{equation}
H_S^i|0\rangle=W_0^i({\boldsymbol \varepsilon};{\boldsymbol \mu},B)|0\rangle, \;\;\;\;
H_S^i|1\rangle=W_1^i({\boldsymbol \varepsilon};{\boldsymbol \mu},B)|1\rangle
\end{equation}
where $W^i({\boldsymbol \varepsilon};{\boldsymbol \mu},B)$ designates the Stark energy for the molecule $i$.

The dipole-dipole coupling term, $V_{d-d}^{i,j}$, between sites $i$ and $j$ is determined by the magnitude and orientation of the dipole moments on those sites, and the lattice spacing of the dipole array,
\begin{equation}
V_{d-d}^{i,j} =\frac{\boldsymbol{\mu}_i\cdot\boldsymbol{\mu}_j-3(\boldsymbol{\mu}_i \cdot {\bf n})(\boldsymbol{\mu}_j \cdot {\bf n})}{|{\bf r}_i-{\bf r}_j|^3}.
\label{coupling}
\end{equation}
Here {\bf n} denotes a unit vector along ${\bf r}_{ij}$.  In the presence of an external field, it becomes appropriate to express $V_{d-d}^{i,j}$ in terms of angles related to the field direction. The result, after averaging over azimuthal angles (for $M = 0$ states are uniformly distributed), reduces to
\begin{equation}
V_{d-d}^{i,j}=\Omega(1-3\text{cos}^2\alpha)\text{cos}\theta_i\text{cos}\theta_j
\label{coupling2}
\end{equation}
where $\Omega=\mu^2/r_{ij}^3$, the angle $\alpha$ is between the ${\bf r}_{ij}$ vector and the field direction and polar angles $\theta_i$ and $\theta_j$ are between the ${\boldsymbol \mu}_i$ and ${\boldsymbol \mu}_j$ dipoles and the field direction. The directional aspect  of the coupling then is governed just by the angle $\alpha$ between ${\bf r}_{ij}$ and the external field direction. We consider the  external field magnitude and direction to be the same for all the polar molecules.

\begin{figure}[ht]
\begin{center}
\includegraphics[width=0.50\textwidth]{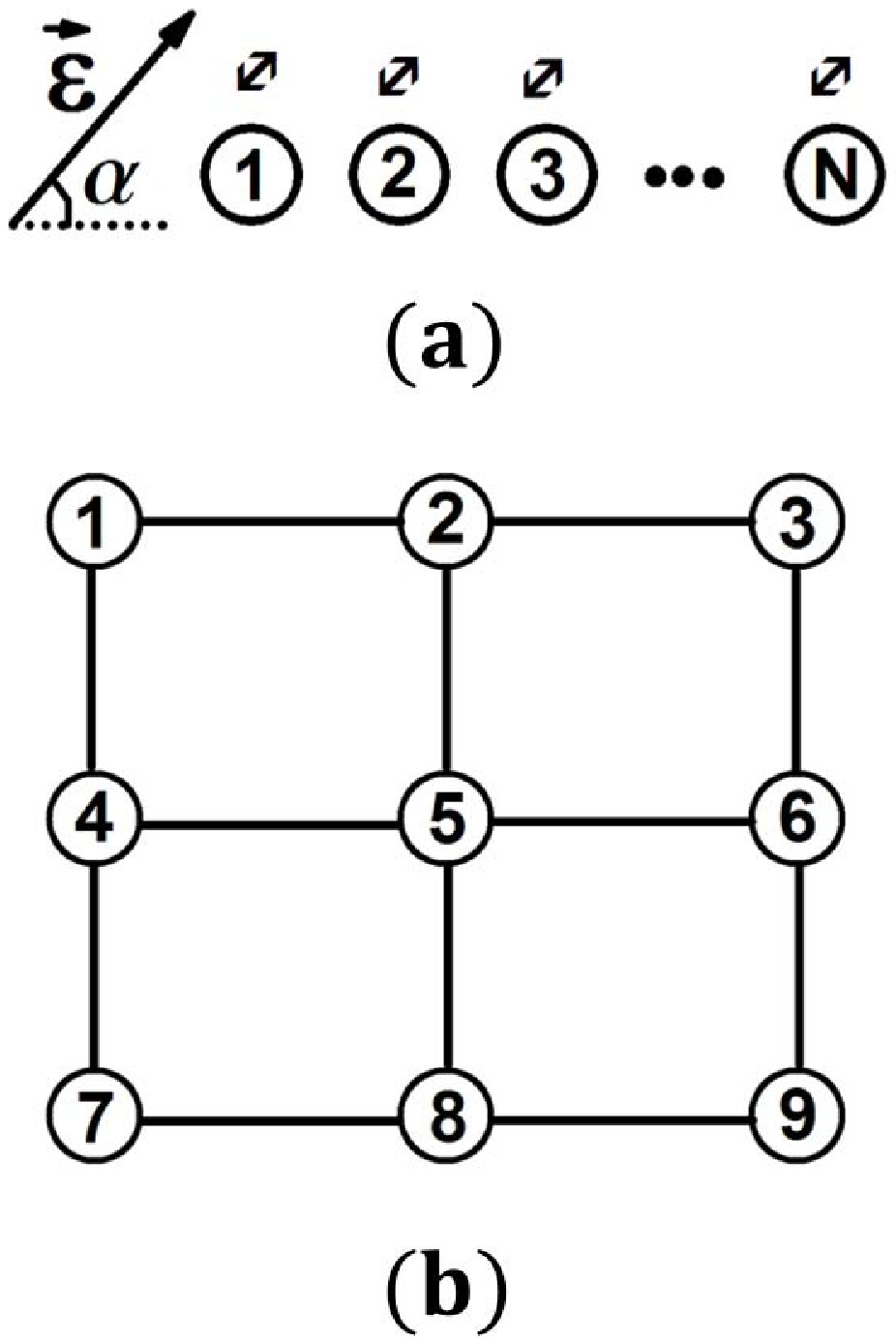}
\end{center}
\caption{Structures of dipole arrays studied: (a) Linear; (b) $3 \times 3$ square lattice. For (b), the external electric field is perpendicular
to the 2-D plane.} \label{coordinate}
\end{figure}

Figure \ref{coordinate} is a schematic depiction of $N$ polar molecules in a (a) linear; (b) square array. The external field is perpendicular to the axis for linear array or the plane of the square array. As our aim is to examine generic behaviors, we can adopt certain simplifying assumptions. For any fixed value of the external field we take all the transition frequencies to be the same, $\triangle W_i=\triangle W=W_1-W_0$, neglecting the small variations required for addressability. We also adopt a standard, representative value of the nearest-neighbor dipole-dipole interaction parameter; for $\alpha = \pi/2$, it is $\Omega = {\bf \mu}^2/|{\bf r}_{i,i\pm 1}|^3$, with $\mu=|\boldsymbol\mu|$ being the permanent body-fixed dipole moment. This idealization avoids specifying the particular effective moments, which in the pertinent range of $\varepsilon=|\boldsymbol\varepsilon|$ typically vary by a factor of up to about three. Any $\Omega_{i,j\neq i}$ can be expressed in terms  of $\Omega$ by merely accounting for the factor $r_{ij}^{-3}$. Thus, the heuristic generic behavior is governed by three variable parameters: the dipole-dipole coupling constant $\Omega$, the intensity of external electric field $\varepsilon$ and the angle $\alpha$. The ranges we considered for $\Omega$ and $\varepsilon$ are: $\Omega/B<10^{-2}$ and  $0<\mu\varepsilon/B<8$. Unless specified otherwise, we set $\alpha = \pi/2$.

\section{Qubit initialization}

The primary goal of initialization is to place the system in a product state such as $|00\cdots 0\rangle$ before any quantum computation is performed. This is the desired form of initial state for most quantum algorithms \cite{Book}. In our case, the ground state is extremely close to $|00\cdots0\rangle$. This can be explained by first-order perturbation theory. If we consider the system of polar molecules \emph{without} dipole-dipole interactions as the unperturbed system $H$, then the ground state of $H$ is \emph{exactly} the product state $|00\cdots 0\rangle$. We then consider the dipole-dipole interaction terms $V_{d-d}^{i,j}$ as the perturbation $V$. Within the first-order approximation we obtain the ground state of the total system as
\begin{equation}\label{eq:psi_1}
|\psi\rangle = |00\cdots 0\rangle + (\Omega/B)\sum_{i=1}^N\langle 0|V^{(i)}|1\rangle|i\rangle
\end{equation}
where $V^{(i)}$ is the projection of $V$ onto the subspace where the $i$-th polar molecule makes a transition from $|0\rangle$ to $|1\rangle$, and $|i\rangle$ is the state where all polar molecules are in $|0\rangle$ except the $i$-th one which is in $|1\rangle$. The scaling factor $\Omega$ comes from Equation \eqref{coupling2} and we regard $B$ as a constant. Assume that $|\langle 0|V^{(i)}|1\rangle|$ is bounded from above by some number $\overline{V}$ that is independent of $N$ (since the dipole-dipole interaction is rather local to nearest neighbors, as will be shown below), $\Omega$ (since we have already taken the $\Omega$ factor in Equation \eqref{coupling2} out of the sum) and $B$. Then the probability of finding the ground state in the space orthogonal to $|00\cdots 0\rangle$, which we denote as $\overline{|000...0\rangle}$, is
\begin{equation}\label{eq:p_not0_p}
P_{\overline{|00\cdots 0\rangle}}\le |\Omega/B|^2\sum_{i=1}^N|\overline{V}|^2 = N\cdot|\Omega/B|^2\cdot|\overline{V}|^2,
\end{equation}
Note that $P_{\overline{|000...0\rangle}}+P_{|000...0\rangle}=1$.

\begin{figure}[ht]
\begin{center}
\includegraphics[width=1.0\textwidth]{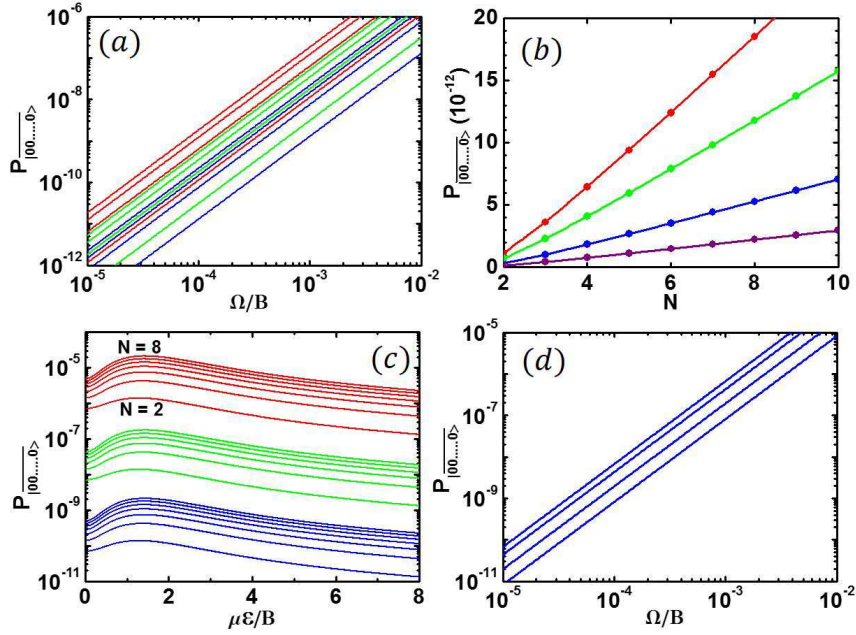}
\end{center}
\caption{(a) $P_{\overline{|000...0\rangle}}$  as a function of $\Omega/B$ for different external electric fields: $\mu\varepsilon/B=$ 2 (red), 3 (green) and 4.9 (blue). For each $\mu\varepsilon/B$, number of polar molecules are from bottom, N = 2, 4, 6, 8, respectively. (b) $P_{\overline{|000...0\rangle}}$  as a function of number of polar molecules in linear array when $\Omega/B = 10^{-5}$ for different external electric fields: $\mu\varepsilon/B=$ 2 (red), 3 (green), 4.9 (blue) and 8 (purple). (c) $P_{\overline{|000...0\rangle}}$  as a function of external electric fields: $\mu\varepsilon/B$  for different $\Omega/B = 10^{-2}$ (red), $10^{-3}$ (green) and $10^{-4}$ (blue). For each $\Omega/B$, number of polar molecules are from bottom, N = 2, 3, 4, ..., 8, respectively. (d) $P_{\overline{|000...0\rangle}}$  as a function of $\Omega/B$ for 2-dimensional arrays in $3 \times 3$ square lattice. External electric fields are from bottom, $\mu\varepsilon/B=$ 2, 3, 4.9 and 8, respectively.}
\label{ground-state}
\end{figure}

Expanding the Hamiltonian in qubit basis states for a system with $N$ qubits, we can get a $2^N\times2^N$ Hamiltonian matrix. Eigenstates can be obtained by diagonalizing the matrix. Figure \ref{ground-state}(a) shows the probability for the ground state to be in state of $\overline{|000...0\rangle}$ as a function of $\Omega/B$ for a linear dipole array with different number $N$ of polar molecules. From Figure \ref{ground-state}(a), one can observe that $P_{\overline{|000...0\rangle}} \propto (\Omega/B)^2$ for the range of $\Omega/B$ that we consider, which is consistant with Equation (\ref{eq:p_not0_p}). Figure \ref{ground-state}(b) shows a linear relationship between $P_{\overline{|000...0\rangle}}$ and the number of polar molecules in a linear array, which is also consistant with Equation (\ref{eq:p_not0_p}). Figure \ref{ground-state}(c) shows how $P_{\overline{|000...0\rangle}}$ changes with external electric field. We fit the data in Figure \ref{ground-state}(a - c) with an empirical formula for $P_{\overline{|000...0\rangle}}$ as a function of the number of polar molecules $N$ as well as other variables of interest:
\begin{equation}
 P_{\overline{|000...0\rangle}} \approx (N-2)\cdot f(\mu \varepsilon/B) \cdot(10000\Omega/B)^2
 \label{fit1}
\end{equation}
where $f(x)$ has the form
\begin{equation}
f(x) = y_0 + \frac{A}{1+\text{exp}[-(x-x_c)/{\triangle x_1}]} \cdot \left\{1- \frac{1}{1+\text{exp}[-(x-x_c)/{\triangle x_2}]} \right\}.
\label{fit2}
\end{equation}
Parameters of Equation (\ref{fit2}) are listed in Table 1. Within acceptable margin of error ($<$ 8\%), Equation (\ref{fit1}) is valid for $0\leq\Omega/B \leq 0.1$, $0 < \mu\varepsilon/B \leq 8$ and $N>3$. 

\begin{table}[!ht]
\begin{center}
\caption{Values of the parameters for $f(x)$ in Equation (\ref{fit2}). }
\begin{tabular}{ccc}
\hline\hline
Parameters& Values& Standard Error \\
\hline
 $y_0$ &3.25724$\times 10^{-11}$  & 4.609$\times 10^{-13}$  \\
 $A$ &8.89294$\times 10^{-10}$ & 3.258$\times 10^{-12}$ \\
 $x_c$&0.78549 & 0.00624 \\
 $\triangle x_1$ &0.28914&0.00214 \\
 $\triangle x_2$  & 1.50288 &0.0078\\
\hline
\end{tabular}
\end{center}
\label{tableC1}
\end{table}

Under conditions envisaged in the proposed designs \cite{Demille,andre,yelin,carr,Book2009,Friedrich}, the dipole-dipole coupling is weak ($\Omega/B$  typically of order $10^{-6}$ to $10^{-4}$). Suppose the nearest neighbor dipole-dipole coupling is $\Omega/B=10^{-4}$ and external field is $\mu\varepsilon/B=2$, then by Equation (\ref{fit1}) for a linear dipole array with $N=1000$ polar molecules, $P_{\overline{|000...0\rangle}} = 3 \times 10^{-7}$. So the ground state system is indeed in pure qubit basis state $|000...0\rangle$. The same conclusion can be drawn for 2-dimensional arrays (see Figure \ref{ground-state}(d)).  So in order to initialize to pure qubit basis state $|000...0\rangle$, we only need to put the system into the ground state, which occurs naturally when we allow the system to reach thermal equilibrium at a sufficiently low temperature. 

\begin{figure}[ht]
\begin{center}
\includegraphics[width=0.7\textwidth]{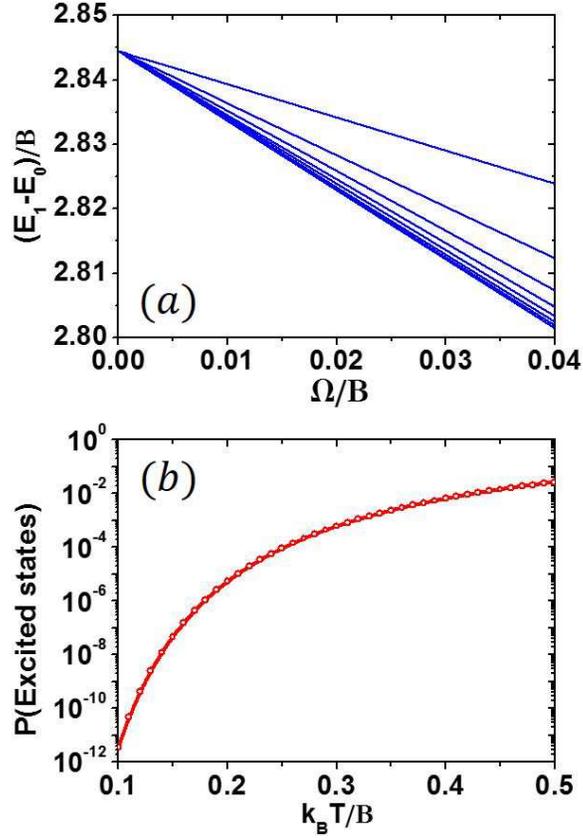}
\end{center}
\caption{ (a) Energy gap between ground state and first excited state as a function of $\Omega/B$ for a linear array with, from top, 2, 3, ..., 9 polar molecules respectively. The external electric field is fixed as $\mu\varepsilon/B = 2$. (b) Probability of thermal excitation to excited states as a function of temperature for a linear array with 8 polar molecules when $\mu\varepsilon/B = 2$ and $\Omega/B=10^{-4}$.} \label{Thermal}
\end{figure}

Figure \ref{Thermal}(a) displays the energy gap between the ground and first excited state $E_1-E_0$ as a function  of $\Omega/B$ up to 0.04 for a linear array in a fixed external field $\mu\varepsilon/B=2$. There is a linear relationship between the energy gap and the dipole-dipole coupling. Using similar arguments that lead to Equations \eqref{eq:psi_1} and \eqref{eq:p_not0_p} we can explain this linear dependence by making use of first-order perturbation theory. In practice, $\Omega/B\leq 10^{-4}$, $E_1-E_0$ is basically the Stark energy difference between $|0\rangle$ and $|1\rangle$ for one qubit ($\triangle W=W_1-W_0$). Adding more polar molecules to the array will not change the energy gap significantly. This is easy to explain physically. Without the dipole-dipole interaction term, the Hamiltonian of the system is diagonal and the eigenstates are qubit basis states \cite{Wei2}. The ground state corresponds to all qubits being in the $|0\rangle$ state and the first excited state corresponds to one of the qubits being in the $|1\rangle$ state. Then the energy gap between ground and first excited state is exactly $\triangle W$. Now, upon introducing the dipole-dipole interaction  term into the Hamiltonian, it is found to be so small compared with the Stark energies that it can be treated as a tiny perturbation that will not change much the eigenenergies.  Therefore, no matter how many polar molecules there are in the array, the energy gap between the ground and first excited state is close to $\triangle W$, which is a function of external  electric field (see Figure \ref{Figure1}). The situation is the same for 2-dimensional dipole arrays. Figure \ref{Thermal} (b) shows the probability for a linear array of $N=8$ polar molecules to be thermally excited to excited states as a function of temperature when $\Omega/B=10^{-4}$ and $\mu\varepsilon/B=2$. Take SrO ($\mu=8.9 D$, $B=0.33 \text{ cm}^{-1}$) as an example, at the proposed temperature of 1 mK \cite{Demille}, for which $k_BT/B=0.002$, the probability for the system to be in an excited state is lower than $10^{-12}$. 

\section{Entanglement measured by Concurrence}

We will deal with the entanglement of formation, $\mathfrak{E}(\rho)$, which characterizes the amount of entanglement needed in order to prepare a state described by a density matrix, $\rho$. (Henceforth, we term $\mathfrak{E}(\rho)$ as just ``entanglement,'' for brevity.) Wootters \cite{Wootters,Hill} has shown that $\mathfrak{E}(\rho)$ for a general state of two qubits can be
quantified by the pairwise $\it concurrence$, $C(\rho)$, which ranges between zero and unity. The relation can be written as \cite{newadd}
\begin{equation}
\mathfrak{E}(\rho) = \xi(C(\rho))
\end{equation}
where $\xi$ is given by
\begin{equation}
\xi(C) = h\left ( \frac{1+\sqrt{1-C^2}}{2} \right )
\end{equation}
with $h(x)=-x \text{log}_2 x - (1 - x) \text{log}_2(1- x)$. The function $\xi(C)$ increases monotonically between zero and unity as $C$ varies from zero to one. The concurrence is given by
\begin{equation}
C(\rho) = \max \{0,\lambda_1-\lambda_2-\lambda_3-\lambda_4 \}
\label{concurrence}
\end{equation}
where the $\lambda_i$'s are the square roots of the eigenvalues, in decreasing order, of the non-Hermitian matrix $\rho\tilde{\rho}$, where $\tilde{\rho}$ is the density matrix of the spin-flipped state, defined as
\begin{equation}
\tilde{\rho} =
(\sigma_y\otimes\sigma_y)\rho^*(\sigma_y\otimes\sigma_y)
\label{rhotilde}
\end{equation}
with $\rho^*$ the complex conjugate of $\rho$; the density matrix is taken in the standard basis which, for a pair of two-level particles, comprises the state vectors $\{|00\rangle,|01\rangle,|10\rangle,|11\rangle\}$. For a system with $N$ qubits, the density matrix for any quantum state is $2^N\times2^N$. The evaluation of pairwise concrrence between qubit A and B only needs the $4\times4$ reduced density matrix as mentioned above. The reduced density matrix can be obtained by ``tracing out'' the rest of the system except the subsystem composed only by qubit A and B \cite{trace}. 

\begin{figure}[ht]
\begin{center}
\includegraphics[width=0.7\textwidth]{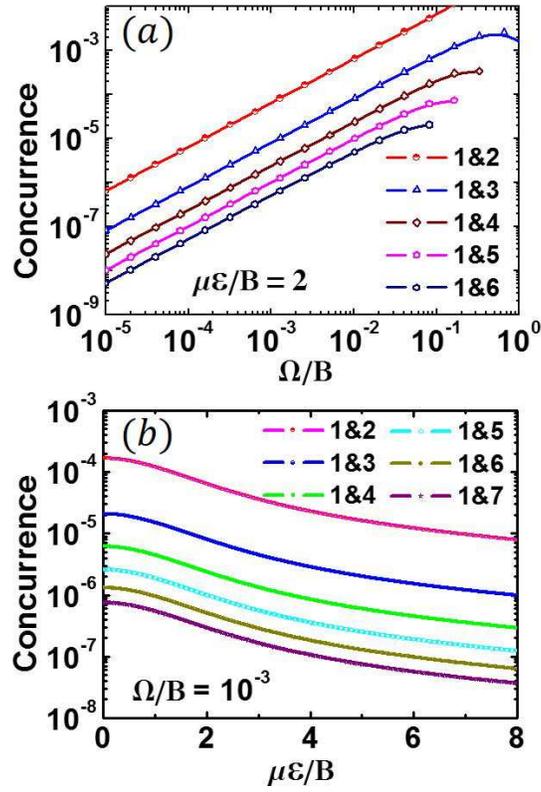}
\end{center}
\caption{ (a) Ground state pairwise concurrences for 1-D dipole arrays ({\it cf.} Figure \ref{coordinate}(a)) with 9 polar molecules as a function
of (a) dipole-dipole coupling constant when $\mu\varepsilon/B=$ 2 and (b) external electric field when $\Omega/B = 10^{-3}$.} \label{areaLaw-linear}
\end{figure}

\begin{figure}[ht]
\begin{center}
\includegraphics[width=0.7\textwidth]{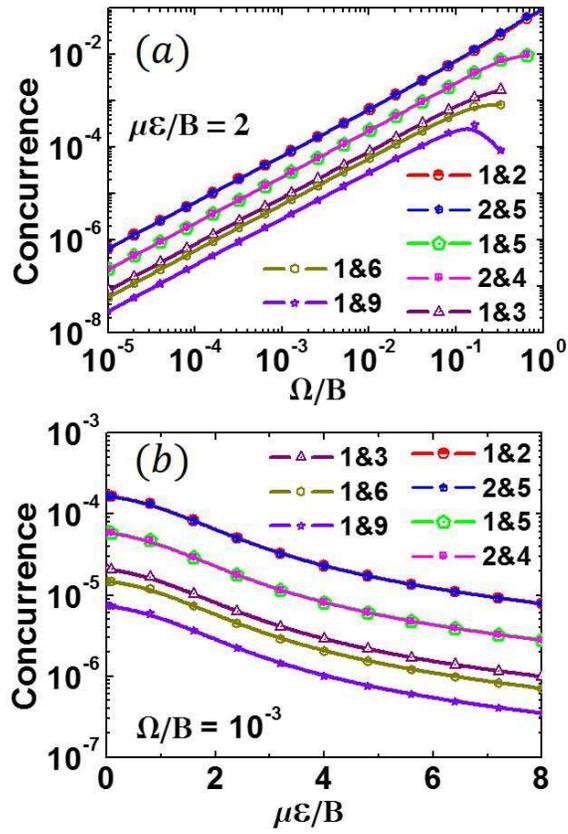}
\end{center}
\caption{ The same as Figure \ref{areaLaw-linear} but for $3\times3$ square dipole arrays ({\it cf.} Figure \ref{coordinate}(b)).} \label{areaLaw-square}
\end{figure}

Figure \ref{areaLaw-linear}(a) shows ground state pairwise concurrence for linear arrays ({\it cf.} Figure \ref{coordinate}(a)) with $N=9$ polar molecules as a function of the dipole-dipole interaction for a fixed external field $\mu\varepsilon/B=$ 2. Figure \ref{areaLaw-linear}(b) shows the same but as a function of the external electric field  for $\Omega/B = 10^{-3}$. Both (a) and (b) exhibit the dominance of entanglement between neighbors for a linear array. The concurences between next nearest neighbors are almost an order of magnitude smaller and the concurences keep decreasing when the distances increase. This means the system can be simplified by limiting our considerations to only next-neighbor interactions. Figure \ref{areaLaw-square} displays the same as Figure \ref{areaLaw-linear} but for a $3\times3$ square array. Entanglement for 2-dimensional arrays are more complicated than 1-dimensional arrays. Interactions between next nearest neighbors can no longer be neglected and the concurrences are only about three-times smaller than those between nearest neighbors. 

The straight lines on the logarithmic coordinates with unit slopes  in Figure \ref{areaLaw-linear}(a) and Figure \ref{areaLaw-square}(a) imply a linear relationship between concurrence and the dipole-dipole coupling constant $\Omega/B$. From earlier arguments this also implies the regime of first-order perturbation theory. Specifically, we have 
\begin{equation}
C_{ij} =K(x)[\Omega_{ij}/B]
\label{Cij-fit}
\end{equation}
where the proportionality factor $K(x)$ is a function of $x=\mu\varepsilon/B$ given in Reference \cite{Wei2}. Equation (\ref{Cij-fit}) is a more generalized version of our previous work for only two polar molecules in pendular states \cite{Wei2}. Equation (\ref{Cij-fit}) holds for any pair of polar molecules in 1-dimensional or 2-dimensional arrays when $\Omega_{ij}/B$ is small ($<0.04$). In \cite{Wei2} we described a numerical analysis that provided an accurate approximate formula:
\begin{equation}
K(x) = A_1 + \frac{A_2}{1+\text{exp}[(x-x_0)/\triangle x]}.
\label{kc}
\end{equation}
The values of the four parameters are listed in Table 2.

\begin{table}[!ht]
\begin{center}
\caption{Values of the parameters for Eq.\ (\ref{kc}).}
\begin{tabular}{ccc}
\hline\hline 
Parameters &  Values &  Standard Error \\
\hline
 $A_1$&0.01092& 0.00015 \\
 $A_2$ &0.2195 & 0.0031 \\
 $x_0$ & 0.9658  & 0.0255 \\
 $\triangle x$  & 0.9743  &  0.0153  \\
\hline
\end{tabular}
\end{center}
\label{table2}
\end{table}

\section{Quantum computing with polar molecules}

Here we discuss potential realizations of various models of quantum computing using arrays of polar molecules in pendular states. In particular, we consider the gate model \cite{Deutsch89,NC11}, measurement-based model \cite{RB01,BR01,RBB03}, instantaneous quantum polynomial-time circuits \cite{Bremner459,Shepherdrspa.2008.0443} and the adiabatic model \cite{FGGS00,kadowaki1998quantum}.

\subsection{Gate model}

\begin{figure}[ht]
\begin{center}
\includegraphics[width=0.8\textwidth]{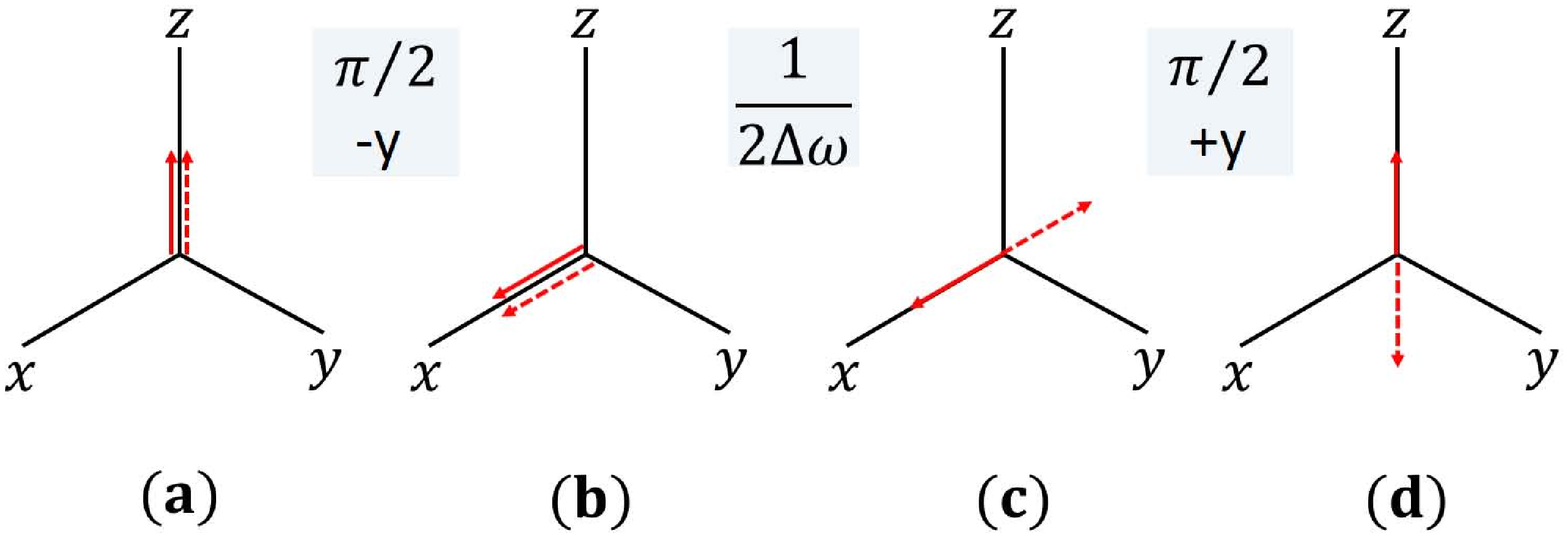}
\end{center}
\caption{The dynamics of target qubit during CNOT operation. The orientation of the red arrow repesents the state of target qubit: up for $|1\rangle$; down for $|0\rangle$; in $xy$ plane for superposition $(|0\rangle+e^{i\Theta}|1\rangle)/\sqrt2$. (a) The target qubit is in state $|1\rangle$: solid for control qubit in state $|0\rangle$ (case 1); dashed for control qubit in state $|1\rangle$ (case 2). (b) After applying $\pi/2$ pulse resonant with both $|00\rangle\leftrightarrow|01\rangle$ (case 1) and $|10\rangle\leftrightarrow|11\rangle$ (case 2) transitions along $-y$ axis, target qubit is at state $(|0\rangle+|1\rangle)/\sqrt2$. (c) After a waiting time of $1/(2\triangle\omega)$, target qubit for the two different cases will have a phase difference of $\pi$. (d) After applying the same pulse as in (b) but along $+y$ axis, the target qubit for case 1 will return to state $|1\rangle$ and for case 2 will reset to $|0\rangle$.} \label{CNOT}
\end{figure}

It is well known that the ability to implement CNOT gates and certain single qubit rotations suffices for universal quantum computation \cite{NC11}.  As for single qubit rotations, the implementation is analogous to that in NMR provided that each polar molecule can be addressed individually. That requires the applied electric field to differ from molecule to molecule such that the $|0\rangle\leftrightarrow|1\rangle$ transition frequency for each site is distinguishable. For one dimensional array, this can be achieved by applying an electric field with appreciable gradient along the array \cite{Demille}. This can also be accomplished by using a trap with micro-electronics to give separate electrodes located under the trapping sites (such devices are
in development for ion traps) \cite{Wei2}. Another alternative way is to use a homogeneous electric field but different polar molecule for each site and every polar molecule should have a unique transition frequency that can be distinguished from the others. The latter two methods work well particularly for 2-dimensional arrays ({\it cf.} Figure \ref{coordinate}(b)). There are two methods to realize CNOT gates with polar molecules.  One is based on multi-target optimal control theory and the scheme is outlined in \cite{Zhu}. The other is the same as that used in NMR \cite{Jones,Cory}. Both methods rely on the frequency shift $\triangle\omega$ induced by dipole-dipole interaction between control and target qubits \cite{Wei2}. The minimum CNOT gate operation times are $10/\triangle\omega$ and $1/(2\triangle\omega)$, respectively \cite{Zhu,Jones}. The latter method is preferred because it is 20 times faster than the first one. The scheme of the second method is outlined in Figure \ref{CNOT}. 

\subsection{Measurement-based model}

One of the potentially realizable gate model processes on our polar molecule system is the preparation of cluster states for measurement-based quantum computation (MBQC) \cite{RBB03}, also known as \emph{one-way quantum computing} \cite{RB01}. The standard procedure for MBQC starts with preparing a specific form of quantum state called \emph{cluster state} \cite{BR01} where the interactions between qubits follow a particular graph (e.g.\ a square lattice). The state preparation can be accomplished by first initializing the system where all qubits are in the $|+\rangle$ state, which is defined as $\frac{1}{\sqrt{2}}(|0\rangle+|1\rangle)$, and then apply controlled-Z on a pair of qubits whenever there is an edge between them in the graph $G$ \cite{NielsenCluster}. Here the Pauli-Z gate acts on a single qubit, it equals to a rotation around the z-axis of the Bloch sphere by $\pi$ radians and the controlled-Z acts on 2 qubits and perform the Z-gate on the second qubit only when the first qubit is $|1\rangle$, otherwise leave it unchanged. After the cluster state is prepared, the computation then proceeds by performing a series of single-qubit measurements. On the polar molecule platform that we consider here, the cluster state can be prepared by initializing the system of molecules at $|0\rangle$ at sufficiently low temperature and applying Hadamard gate to each molecules, resulting in an all-$|+\rangle$ state. Here the Hadamard gate acts on a single qubit, it maps the basis state $|0\rangle$ to $(|0\rangle+|1\rangle)/\sqrt2$ and $|1\rangle$ to $(|0\rangle-|1\rangle)/\sqrt2$. The controlled-Z gates could then be applied to yield the cluster state needed. Since the controlled-Z gates are only applied to qubits that are spatially adjacent in the graph, the implementation is greatly simplified compared with general quantum circuits where a pair of qubits arbitrarily far apart may need to be entangled. We note that in the literature there have been experimental realizations of cluster states using linear optics \cite{WRRS+05,LZCG+07}, cavity QED \cite{PhysRevA.77.062312}, neutral atoms \cite{PhysRevA.85.012328}, trapped ions \cite{PhysRevA.79.052324,PhysRevLett.102.170501}, and atomic ensembles \cite{PhysRevA.79.022304}.

\subsection{Instantaneous quantum polynomial time circuits}

Although, as we have mentioned previously, it is feasible to realize universal quantum computing with our polar molecule system, there are non-universal classes of quantum circuits that produce output distributions which are believed to be hard to sample from using randomized classical algorithms. In other words, one does not need to push for universal quantum computation to construct quantum processes that are classically hard to simulate. Of course, the belief about such difficulty with classical computers is based on well-known conjectures in computational complexity. Here we consider a class of circuits known as IQP (Instantaneous Quantum Polynomial time), which was introduced in Refs.\ \cite{Bremner459,Shepherdrspa.2008.0443}. An $n$-qubit IQP circuit takes the form $H^{\otimes n}DH^{\otimes n}$ where $D$ is a diagonal unitary operator. Equivalently one could regard an IQP circuit as a sequence of commuting quantum gates. As simple as they seem, the output distributions of IQP circuits are classically hard to sample from in the worst case. For an arbitrary $D$ of poly$(n)$ gates, it is $\# P$-hard \cite{BMS15,GG14,S85} to compute for instance the probability $p=|\langle 00\cdots 0|H^{\otimes n}DH^{\otimes n}|00\cdots 0\rangle|^2$. However, an IQP circuit is relatively simple to construct using our polar molecule setup. The steps of Hadamard transforms, $H^{\otimes n}$, can be realized in parallel. The diagonal unitary $D$ can also be readily realized with explicit circuit constructions \cite{Childs04,WGMA14}. In particular, we stress that for any $n$-qubit diagonal unitary $D$, using techniques from \cite{WGMA14} we could construct an approximation circuit $\hat D$ acting on the same $n$ qubits using only nearest neighbor CNOT gates (see Appendix \ref{sec:diagonal}). 

For instance, recently it was shown by Qiang \emph{et al.\ }\cite{QLMA+16} that one could gain an exponential speedup in simulating continuous time quantum walk on circulant graphs (graphs whose adjacency matrices satisfy the property where the row $j+1$ can be obtained by rotating row $j$ by one element) compared with the best classical algorithm. It is known that any Hamiltonian $H$ for a continuous time quantum walk on any circulant graph can be diagonalized by the unitary Fourier Transform \cite{Gray01toeplitzand}: $H=Q^\dagger \Lambda Q$ where $\Lambda$ is diagonal. The evolution under $H$ then becomes $e^{-iHt}=Q^\dagger DQ$ where $D=e^{-i\Lambda t}$ simulates a diagonal Hamiltonian. Quantum Fourier transform $Q$ is well-known to be efficiently realizable, and general schemes also exist for simulating diagonal unitaries \cite{Childs04,WGMA14}. Hence a quantum walk on a circulant graph of $2^n$ nodes is efficiently realizable in the gate model with $n$ qubits. More interesting is the prospect of replacing the $n$-qubit quantum Fourier transform $Q$ with easier-to-implement Hadamard gates $H^{\otimes n}$ when it comes to computing the probability of obtaining $|00\cdots 0\rangle$ in the final state \cite{QLMA+16}
\begin{equation}
\begin{array}{ccl}
p_{00\cdots 0}  & = & |\langle 00\cdots 0|Q e^{-i\Lambda t} Q^\dagger|00\cdots 0\rangle|^2 \\[0.1in]
 & = & |\langle 00\cdots 0|H^{\otimes n}e^{-i\Lambda t} H^{\otimes n}|00\cdots 0\rangle|^2.
\end{array}
\end{equation}
Evidence presented in \cite{QLMA+16} for the difficulty of computing $p_{00\cdots 0}$ on a classical computer is also based on the notion of instantaneous quantum polynomial time (IQP).

\subsection{Adiabatic quantum computing}

There has already been rather impressive implementation of adiabatic quantum computing at a scale of at least hundreds of spins \cite{BRIW+14,SQVL14,VMBR+14,VMKO15,HJAR+15,KHZO+15}. However, one of the main issues faced by such implementation is control of precision, \emph{i.e.\ }the dynamic range of field values which a device must be able to resolve in order to embed the intended eigenspectrum to a desired accuracy. Current benchmark of the D-Wave 2X system has reached control precision of maximum $2*127=254$ different values \cite{KYNH+15}. We argue that a potential implementation of adiabatic evolution on the polar molecules could yield a dynamic range on the order of $10^4$ in interaction strength. The basic idea is to deviate from our earlier proposal on gate model by identifying the state $|J=1,m=0\rangle$ as the computational $|0\rangle$ state and $|J=1,m=1\rangle$ as the computational $|1\rangle$ state. Then the energy gap between the two states could be tuned arbitrarily from $0$ to $2.5B$ (see Figure \ref{Figure1}). Consider an initial state of the polar molecules under a strong electric field. Under a strong field the state of the molecular system must be oriented along the field. Then we gradually weaken the field and vary its shape to certain prescribed distribution between the molecules such that the ground state under the new distribution is hard to find. A major advantage of our setup compared with D-Wave is the wide range of tunable interactions. The energy gap between $|0\rangle$ and $|1\rangle$ could be tuned by the strength of the electric field from 0 to $10^4$ times as large as $V_{d-d}$, which is far beyond D-Wave's current capability.  One concern is whether during the adiabatic procedure of weakening the electric field the entire system of polar molecules could undergo a phase transition. For the purpose of adiabatic computation one requires that an energy gap be constantly maintained during the adiabatic evolution, which is a property that needs to be checked.

\section{Acknowledgment}

Partial funding for this research was provided by Qatar National Research Foundation (QNRF), grant no. NPRP 7-317-1-055. We would also like to acknowledge Shanghai Natural Science Foundation, grant no. 15ZR1411300.

\section{Keywords}

polar molecule, pendular state, quantum computer

%%%%%%%%%%%%%%%%%%%%%%%%%%%%%%%%%%%%%%%%%%%%%%%%%%%%%%%%%

%alternatively you may use

%\bibliographystyle{unsrt}
%\bibliography{mymanuscript.bib}

%and send the .bib file along with your manuscript

%%%%%%%%%%%%%%%%%%%%%%%%%%%%%%%%%%%%%%%%%%%%%%%%%%%%%%%%%

\appendix
\section{Quantum algorithm for arbitrary diagonal unitary}\label{sec:diagonal}

Here we present a method for realizing arbitrary diagonal unitary $D$ using nearest-neighbor CNOT gates.
The circuit $\hat D$ consists of $O(\frac{1}{\epsilon}\log^2\frac{1}{\epsilon})$ single-qubit and \emph{nearest-neighbor} CNOT gates and $\|D-\hat D\|\le\epsilon$. Our argument is as follows: 
\begin{enumerate}
\item
$\hat D$ is a sequence of unitaries $\hat U_j$ such that $\hat D = \hat U_1\hat U_2\cdots \hat U_k$ where $k=O(1/\epsilon)$.
\item
According to the scheme in \cite{WGMA14}, each $\hat U_j$ is realized by a single-qubit rotation gate on some qubit, combined with CNOT gates targeted on the qubit corresponding to the most significant non-zero bit (MSB) in the binary expansion of the integer $j$ and controlled on the qubits corresponding to the 1's other than the MSB. Since $j\le k=O(1/\epsilon)$, each CNOT acts on two qubits that are at most $O(\log 1/\epsilon)$ apart. For three qubits $A,B,C$, we have the identities 
\begin{equation}\label{eq:cnot}
\begin{array}{ccl}
\text{CNOT}_{AC} & = & \text{SWAP}_{AB}\text{CNOT}_{BC}\text{SWAP}_{AB} \\
\text{SWAP}_{AB} & = & \text{CNOT}_{AB}\text{CNOT}_{BA}\text{CNOT}_{AB}
\end{array}
\end{equation}
where $\text{CNOT}_{ij}$ stands for a CNOT gate with $i$ as the control and $j$ as the target bit and $\text{SWAP}_{ij}$ swaps the qubits $i$ and $j$. We could use the identities in \eqref{eq:cnot} to decompoe a CNOT gate acting on positions that are $m$ qubits apart into $O(m)$ CNOT gates acting only on nearest-neighbor qubits. Hence each CNOT gate in $\hat U_j$ can be realized by $O(\log 1/\epsilon)$ nearest-neighbor CNOT gates. The  binary expansion of $j$ contains at most $O(\log 1/\epsilon)$ bits, implying that each $\hat U_j$ can be realized by $O(\log^21/\epsilon)$ nearest-neighbor CNOT gates.
\item Putting everything together, $\hat D$ can be realized using $O(1/\epsilon)$ single qubit rotations and $O(\frac{1}{\epsilon}\log^2\frac{1}{\epsilon})$ nearest-neightbor CNOT gates. 
\end{enumerate}

\newpage

\newpage

\newpage

\newpage

\newpage

\newpage

\newpage

\end{document}